\documentclass[aps,prl,twocolumn,showpacs,groupedaddress,floatfix]{revtex4}

\usepackage{graphicx} 
\usepackage{dcolumn}
\usepackage{epsfig} 
\usepackage{amssymb} 
\usepackage{amsmath}  
\usepackage{color} 
\newcommand{\gammap}{\dot{\gamma}}
\newcommand{\gammapeff}{\gammap_{\hbox{\rm\scriptsize true}}}
\newcommand{\mms}{~mm\,s$^{-1}$}

\begin{document}

\title{Spatio-temporal dynamics of wormlike micelles under shear}

\author{Lydiane B\'ecu\footnote{Corresponding author: {\tt becu@crpp-bordeaux.cnrs.fr}}}
\affiliation{Centre de Recherche Paul Pascal, Avenue Schweitzer, 33600 Pessac, FRANCE}
\author{S\'ebastien Manneville}
\affiliation{Centre de Recherche Paul Pascal, Avenue Schweitzer, 33600 Pessac, FRANCE} 
\author{Annie Colin\footnote{Present address: Rhodia Laboratoire du Futur, CNRS-FRE 2771, B\^{a}t. B de l'Institut
Europ\'een de Chimie et Biologie, 2 rue Robert Escarpit, 33607 Pessac Cedex.}}
\affiliation{Centre de Recherche Paul Pascal, Avenue Schweitzer, 33600 Pessac, FRANCE}

\date{\today}
\begin{abstract}Velocity profiles in a wormlike micelle solution (CTAB in D$_2$O) are recorded 
using ultrasound every 2~s
after a step-like shear rate into
the shear-banding regime. 
The stress relaxation occurs over more than six hours and corresponds to
the very slow nucleation and growth of the high-shear band. 
Moreover, oscillations of the interface position with a period
of about 50~s are observed during the growth process. 
Strong wall slip, metastable states and transient nucleation of three-band flows
are also reported and discussed in light of previous experiments and theoretical models.
\end{abstract}
\pacs{83.60.-a, 83.80.Qr, 47.50.+d, 43.58.+z}
\maketitle

Wormlike micelle solutions exhibit strong shear-thinning behavior
due to the coupling between their  microstructure and the 
flow. Along the steady-state flow curve (shear stress $\sigma$~vs. shear rate $\gammap$),
 a drop of up to three orders of magnitude in the effective viscosity $\eta=\sigma/\gammap$
 is observed in a very narrow stress range leading to a stress plateau (for a review, see
 Refs.~\cite{Larson:1999,Edimbourg:2000}). 
Such a sharp transition made wormlike micelle solutions appear
as a model system to study shear-induced effects in complex fluids. 

A partial understanding of the flow curve of wormlike micelles has
emerged thanks to local scattering and velocimetry experiments. 
Above a critical shear rate $\gammap_1$, a birefringent band 
normal to the velocity gradient occupies an increasing part
of the gap as the shear rate is increased \cite{Cappelaere:1995,Lerouge:1998}. 
This shear-induced birefringent band was clearly identified as a nematic phase
 in the case of concentrated solutions and has a low viscosity compared to the isotropic phase
\cite{Cappelaere:1995,Berret:1994a}.
Moreover, the velocity field was shown to separate into
two differently sheared bands \cite{Britton:1997,Salmon:2003c}:
a weakly sheared region flowing at $\gammap_1$
 and a highly sheared region at $\gammap_2$,
the upper limit of the stress plateau. 
In the well-documented system CPCl/NaSal in brine \cite{Rehage:1991}, the equilibrium position
of the interface between the two shear bands
increases linearly with $\gammap$ consistently
with the ``lever rule'' \cite{Salmon:2003c}.
However, such a simple shear-banding scenario
remains controversial due to (i) temporal fluctuations
of the flow field and (ii) slow transients that
question the existence of truly stationary states
\cite{Britton:1997,Mair:1997,Fischer:2000b}.

At this stage, due to the limited temporal resolution of current local velocimetry
techniques (Nuclear Magnetic Resonance \cite{Britton:1997,Mair:1997,Fischer:2000b}
and Dynamic Light Scattering
\cite{Salmon:2003c}), a {\it time-resolved}
description of the velocity field is still missing. In this Letter,
velocity profiles obtained in cylindrical Couette geometry
by high-frequency ultrasonic velocimetry \cite{Manneville:2003pp_a}
and recorded every 2~s {\it simultaneously} to global rheological data
are presented for a startup experiment followed over about six hours.
For one of the much studied micellar systems (CTAB in D$_2$O), we show
that two very different time scales are involved in the dynamics of shear banding.
(i) A three-step nucleation and growth of a high-shear band occurs on a few hours. 
(ii) ``Fast'' oscillations of the band position with a period
of about 50~s are pointed out. 
Subtle additional effects such as strong wall slip and transient nucleation of three-band flows
are also reported and discussed.

\begin{figure}[htbp]
\begin{center}
\scalebox{1}{\includegraphics{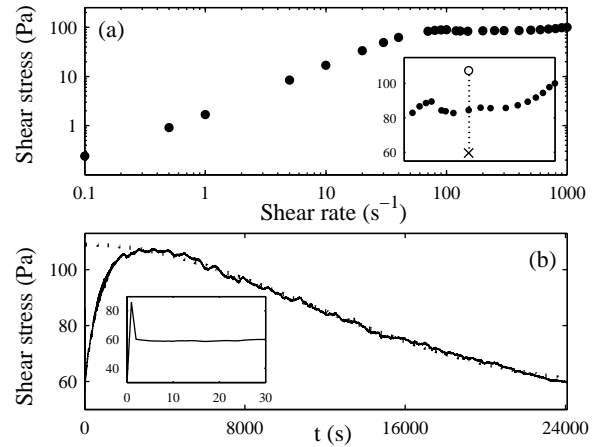}}
\end{center}
\caption{\label{f.rheol}(a) ``Quasistatic" flow curve $\sigma(\gammap)$ obtained at $T=44^\circ$C under controlled shear rate (15 min per data point)
by averaging the shear stress over the last 100~s of each step.
Inset: enlargement over $\gammap=60$--1000 s$^{-1}$ showing the stress plateau. The dotted
line corresponds to the 
stress response to a step-like shear rate from 0 to 200~s$^{-1}$. $\circ$ and $\times$ symbols indicate respectively the maximum and the last value of $\sigma$. (b) Temporal response of the shear stress $\sigma(t)$
after a step-like shear rate from 0 to 200~s$^{-1}$ (thin line). The thick dotted
line is the sigmoidal relaxation with 
$\sigma_\infty=56.4$ Pa, $\sigma_M=109$ Pa, and $\tau_{ss}=15450$~s (see text).
Inset: enlargement over $t=0$--30~s showing the stress overshoot.
}
\end{figure}

We focus on the salt-free wormlike micelle solution made of Cetyl Trimethyl
Ammonium Bromide (CTAB) at 20\% wt. in deuterated water (D$_2$O) at a temperature of $44^\circ$C i.e.
in the vicinity of the isotropic--nematic (I--N) transition that occurs at $T_{IN}=39^\circ$C
\cite{Cappelaere:1995,Fischer:2000b}.
Rheological data are measured using a stress imposed rheometer
 (TA Instruments AR 1000) and a Couette cell of
 inner radius $R_1=24$~mm and gap width $e=1.1$~mm.
The cell is surrounded by a solvent trap containing water to prevent evaporation. 
Our local velocimetry technique is based on time-domain cross-correlation
of high-frequency ultrasonic signals backscattered by the moving fluid. 
Post-processing of acoustic data allows us to record a velocity profile 
in 0.02--2~s with a spatial resolution of 40~$\mu$m
(see Ref.~\cite{Manneville:2003pp_a} for more details).
In order to enhance the scattering properties of our system,
we add a small amount of colloidal particles (1\% wt. polystyrene spheres
of diameter 3 to 10~$\mu$m). We checked that the linear rheological properties
as well as the plateau behavior were not significantly
affected by the addition of those scatterers.

Using the shear rate imposed mode of the rheometer, we first apply increasing shear 
rates during a scanning time of 900~s per step.
The resulting flow curve $\sigma(\gammap)$ shown in Fig.~\ref{f.rheol}(a)
presents a stress plateau at $\sigma^\star\approx 84$~Pa
that extends from $\gammap_1\approx 50$~s$^{-1}$ to $\gammap_2\approx 500$~s$^{-1}$
corresponding to a drop in the effective viscosity by an order of magnitude.
Note that, in the system under study, the features of the plateau
($\sigma^\star$, $\gammap_1$, and $\gammap_2$) depend dramatically on the scanning time
as already reported in Ref.~\cite{Cappelaere:1995}. The bump in $\sigma(\gammap)$ revealed
by the inset of Fig.~\ref{f.rheol}(a) is characteristic of underlying metastable states
\cite{Grand:1997}. Thus, between $\gammap_1$ and $\gammap_2$ and
even for scanning times of 15 min, the flow curve of Fig.~\ref{f.rheol}(a)
does not represent true equilibrium states but rather describes qualitatively the
plateau behavior of our system.

In the following, we apply a step-like shear rate into the plateau region from 0 to 200~s$^{-1}$
at time $t=0$ and 
we record simultaneously the shear stress response $\sigma(t)$ and the velocity profiles
for about six hours.
As seen in Fig.~\ref{f.rheol}(b), after an elastic overshoot during the very first seconds,
$\sigma(t)$ increases from 60 to 105~Pa in 2000~s, then remains nearly constant at 107~Pa for 2500~s,
and finally decreases very slowly for the rest of the experiment.
This ultraslow decrease may be well fitted by a sigmoidal function 
$\sigma(t)=\sigma_\infty+(\sigma_M-\sigma_\infty)\exp((-t/\tau_{ss})^2)$
with $\tau_{ss}=15450$~s and $\sigma_\infty=56.4$~Pa \cite{Grand:1997}. 
Note that such a long equilibration time may be surprising for a wormlike micelle
solution but is consistent with previous rheological data on the same salt-free system 
close to the I--N transition \cite{Cappelaere:1995}. We checked that this very long time scale
cannot be attributed to any evaporation effect by measuring the weight fraction of
the sample before and after shearing using thermogravimetric analysis.

\begin{figure}[htbp]
\begin{center}
\scalebox{1.0}{\includegraphics{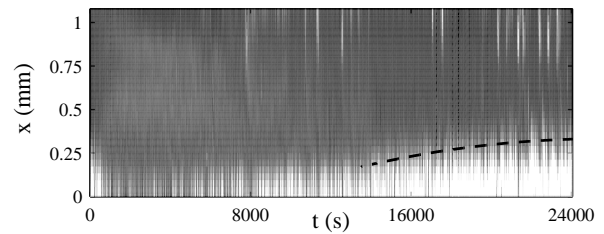}}
\end{center}
\caption{Local shear rate $\gammap(x,t)$. A linear gray scale is used: black and white correspond respectively to $\gammap=0$~s$^{-1}$
and $\gammap\ge 300$~s$^{-1}$.
The thick dashed line is given by
$\delta(t)=\delta(t_0)+C\int_{t_0}^t (\sigma(t')-\sigma_\infty)\,\hbox{\rm d}t'$,
with $t_0=13500$~s, $\delta(t_0)=0.17$~mm, and $C=1.17$\,10$^{-6}$~mm\,Pa$^{-1}$\,s$^{-1}$.
$\sigma_\infty=56.4$~Pa was deduced from Fig.~\ref{f.rheol}(b).}
\label{f.spatiotemp200}
\end{figure}

\begin{figure}[htbp]
\begin{center}
\scalebox{1.0}{\includegraphics{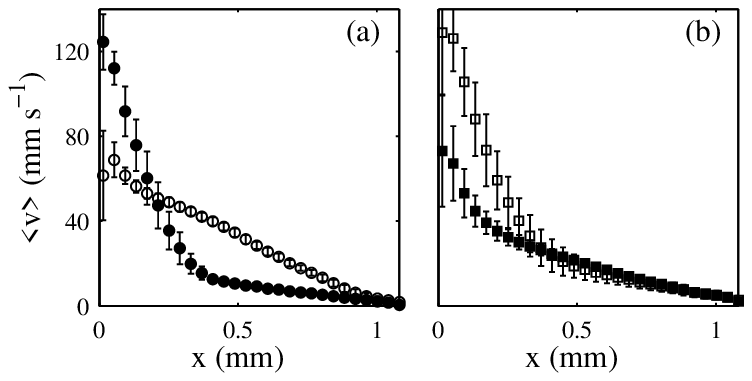}}
\end{center}
\caption{Description of the slow dynamics. Velocity profiles $<v(x)>$
averaged over (a) $t=2$--25 ($\bullet$), 4400--4700 ($\circ$), (b) 12000--12230 ($\blacksquare$), and
23600--23840~s ($\square$). The error bars are the standard deviations of these estimates. The rotor
velocity at $\gammap=200$~s$^{-1}$ is $v_0=196$\mms.}
\label{f.slow}
\end{figure}

Figure~\ref{f.spatiotemp200} gathers the full ultrasonic velocimetry data
on a spatio-temporal diagram of the local shear rate $\gammap(x,t)$ 
calculated from the velocity $v(x,t)$ as $\gammap(x,t)=-(R_{1}+x)\frac{\partial}{\partial x}\,\frac{v(x,t)}{R_{1}+x}\,$.
The abscissae correspond to time $t$ and the ordinates to the radial position $x$ inside
the gap of the Couette cell with $x=0$ ($x=e$) at the rotor (stator).
This representation reveals that the flow field slowly evolves in time but also
undergoes ``fast'' fluctuations \cite{Remark:moviesUSV}.
By first looking at the velocity profiles averaged over at most 6~min, we may distinguish three different regimes
(indicated by arrows on Fig.~\ref{f.spatiotemp200} and summarized in Fig.~\ref{f.slow}) along the slow stress relaxation. 

At the inception of the flow and for about 200~s, {\it two bands} bearing 
different shear rates, $\gammap_{h}\approx 360$~s$^{-1}$ and $\gammap_{l}\approx 16$ s$^{-1}$, coexist in the gap 
of the Couette cell (see $\bullet$ symbols in Fig.~\ref{f.slow}(a)).
The interface between the shear bands is located at $\delta\approx 0.3$~mm from the stator.
Moreover, wall slip is present
since the fluid velocity neither reaches the rotor velocity
$v_0=196$\mms\ at $x=0$
nor perfectly vanishes at $x=e$. 
We define the slip velocities at the rotor $v_{s1}$
and at the stator $v_{s2}$ as the differences between the fluid velocity near the wall and the corresponding wall velocity, 
and get $v_{s1}\approx 75$\mms\ and $v_{s2}\approx 3$\mms.
Slip velocities are thus strongly 
dissymmetric and demonstrate that wall slip occurs preferentially in the low viscosity phase.  

Second, after this first regime, the highly sheared band disappears so that the
bulk flow, when averaged over a few minutes, becomes {\it homogeneous} for about 8000~s.
This regime is reminiscent of the ``milky'' turbid phase observed
at ``intermediate'' times in Ref.~\cite{Lerouge:1998}.
Strong wall slip is detected at the rotor ($v_{s1}\approx 134$\mms) while slip at the stator
remains negligible (see $\circ$ symbols in Fig.~\ref{f.slow}(a)). 
Note that the sliding layer at the rotor
supports a very high shear rate of the order of 3000~s$^{-1}$ and thus
unambiguously differs from a band of nematic phase thinner than our resolution of 40~$\mu$m
(that would flow at $\gammap_h$).
Another striking feature of this homogeneous velocity profile is the value of the ``true'' shear rate
(calculated by removing the contributions of sliding layers) $\gammapeff\approx 63$~s$^{-1}$, which
falls between the local shear rates $\gammap_h$ and $\gammap_l$
measured previously in the high- and low-shear bands. For such a shear rate, the flow is
expected to show (at least) two shear bands.
This suggests that during this part of the experiment, the system explores
a metastable branch of the flow curve located above the stress plateau at $\sigma^\star$ \cite{Grand:1997}.
Indeed, this second stage roughly corresponds to the increase and stagnation of the
stress response $\sigma(t)$ well above both the asymptotic 
value $\sigma_\infty=56.4$~Pa and the ``quasistatic'' plateau value $\sigma^\star\approx 84$~Pa (see Fig.~\ref{f.rheol}).

Finally, from $t\approx 8000$~s until the end of the experiment, the velocity profiles
show the {\it growth} of a low viscosity
layer in the vicinity of the rotor (see Fig.~\ref{f.slow}(b))
qualitatively similar to the ``long'' time behavior reported in flow birefringence
experiments \cite{Lerouge:1998}.
This highly sheared band expands until it fills roughly one third of the gap.
During this growth stage, the slip velocity $v_{s1}$ decreases from 134 to 64\mms\ so that the local
shear rate in the high-shear band $\gammap_h$ remains roughly 
constant and equal to 360~s$^{-1}$.
Moreover, the weakly sheared band bears a constant local shear rate $\gammap_l\approx 30$~s$^{-1}$.
Thus, although the asymptotic state is still not perfectly reached, the interface position
at the end of the experiment is compatible with the lever rule: 
$\delta=(\gammapeff-\gammap_l)/(\gammap_h-\gammap_l)e\approx 0.3$~mm.

\begin{figure}[htbp]
\begin{center}
\scalebox{1.0}{\includegraphics{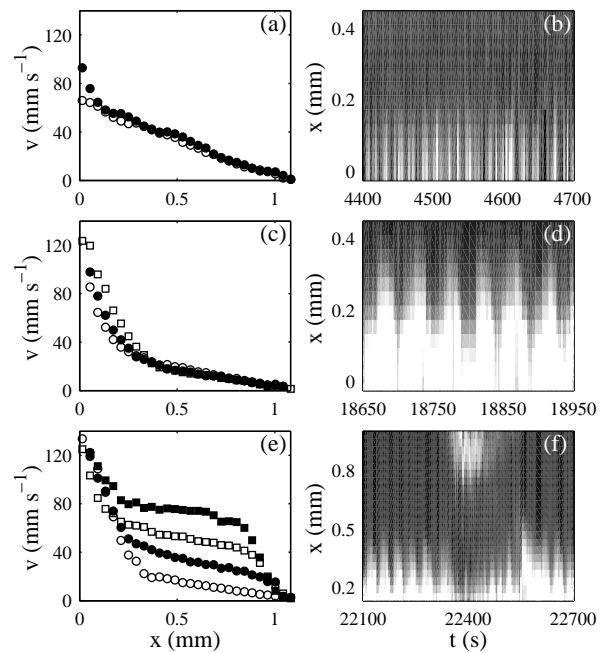}}
\end{center}
\caption{Description of the fast dynamics.
(a) Velocity profiles during the intermittent apparition of a highly sheared band at the rotor ($x\lesssim 150~\mu$m) at $t=4505$ ($\bullet$), 
and 4508~s ($\circ$).
(b) Enlargement of $\gammap(x,t)$ over $t=4400$--4700~s.
(c) Velocity profiles during one interface oscillation at $t=18800$ ($\circ$), 18810 ($\bullet$), and 18818~s ($\square$).
(d) Enlargement of $\gammap(x,t)$ over $t=18650$--18950~s.
(e) Velocity profiles during the nucleation of a second highly sheared band at $t=22292$ 
($\circ$), 22350 ($\bullet$), 22388 ($\square$), and 22463~s ($\blacksquare$).
(f) Enlargement of $\gammap(x,t)$ over $t=22100$--22700~s.}
\label{f.fast}
\end{figure}

\begin{figure}[htbp]
\begin{center}
\scalebox{1.0}{\includegraphics{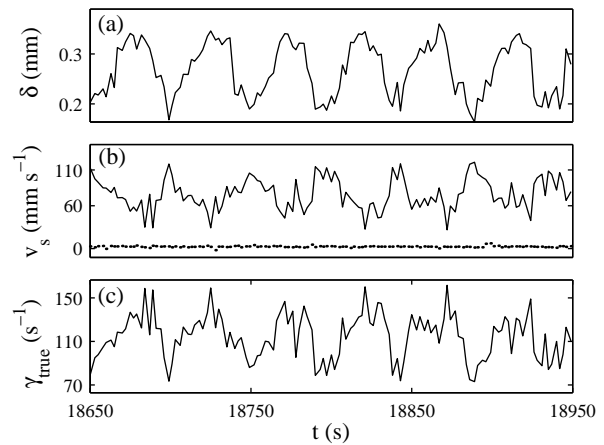}}
\end{center}
\caption{Temporal fluctuations of (a) the position of the interface $\delta(t)$, (b)
the slip velocities at the rotor $v_{s1}(t)$ (solid line) and at the stator $v_{s2}(t)$
(dotted line), and (c) the true shear rate $\gammapeff(t)$, recorded during the
oscillations of Fig.~\ref{f.fast}(d)}
\label{f.fluctu} 
\end{figure}

Let us now get a closer look at the fast dynamics superimposed to the slow evolution
described above.
In the regime where the flow appears homogeneous on average ($t\approx 200$--8000~s),
Fig.~\ref{f.fast}(a) and (b) reveal that a thin band of width
$\delta \lesssim 150~\mu$m is actually nucleated and destroyed within short intervals of a few seconds
unevenly every 5 to 20~s.
When present, the band is sheared at $\gammap_h$ and can be clearly distinguished from wall slip.
Ref.~\cite{Lerouge:1998} also reports the presence of a small shear band at the rotor 
as the ``milky'' phase disappears during the second regime.

Moreover, Fig.~\ref{f.fast}(c) and (d) show that, during the expansion of the
highly sheared band ($t\approx 8000$--24000~s), the interface is in fact subjected to
periodic oscillations with a period of 50~s.
As Fig.~\ref{f.fluctu} points out, the slip velocity $v_{s1}(t)$ oscillates in phase
opposition to the interface position $\delta(t)$, whereas the true shear rate 
$\gammapeff(t)$ is synchronized with $\delta(t)$.
These oscillations sometimes give way to the nucleation and
destruction of a second highly sheared band at the stator (see Fig.~\ref{f.fast}(e) and (f)).
This phenomenon develops within roughly 200~s and takes place rather 
intermittently.
According to Ref.~\cite{Radulescu:1999}, such a transient event is conceivable  in Couette
geometries of low enough curvature.
On the spatio-temporal data $\gammap(x,t)$, it shows up as a white
patch for $x\gtrsim 0.8$~mm. This process seems to occur more frequently at the end of the experiment (see Fig.~\ref{f.spatiotemp200}). 
Oscillations similar to those of Fig.~\ref{f.fast}(d) are clearly visible before and after the event 
of Fig.~\ref{f.fast}(f).

Let us conclude by discussing the main question raised by our study: what is the origin of the various
time scales involved in the spatio-temporal dynamics of our micellar system?
Concerning the {\it slow time scales}, the present results reveal a strong
similarity with the evolution reported in Ref.~\cite{Lerouge:1998}.
However, our system is more than two orders of magnitude slower:
``intermediate'' and ``long'' times correspond respectively
to 200--8000~s and 8000--24000~s here
vs. 5--35~s and 35--175~s in Ref.~\cite{Lerouge:1998}.
Such a huge discrepancy could be explained by the differences in composition,
electrostatic screening, or proximity to the I--N transition.
In the framework of theoretical approaches derived from the Johnson-Segalman model
\cite{Radulescu:1999}, such long time scales mean that the diffusion coefficient $D$ of the
shear stress through the interface is vanishingly small. The interface speed
is predicted to scale as $\sqrt{D}(\sigma-\sigma_\infty)$. The dashed line in 
Fig.~\ref{f.spatiotemp200} shows that
our data are fairly compatible with this prediction. 

Finally, the presence of {\it fast time scales} ranging from 5 to 50~s and
thus strongly exceeding the micelle relaxation time $\tau_R\approx 40$~ms remains very puzzling
\cite{Remark:tempo}.
Recent phenomenological models including dynamical equations for the micelle
length and/or relaxation time predict oscillations and even chaotic-like states of the flow field
that could be connected to the dynamics observed experimentally \cite{Fielding:2003pp}.
However, we believe that {\it wall slip dynamics} may play a major role in 
our experiments. Indeed, time scales of the same order have been found
after small shear rate jumps between two banded states \cite{Radulescu:2003}. 
They were related to the interface reconstruction after band destabilization.
Here, due to wall slip, the true shear rate
is not stationary (see Fig.~\ref{f.fluctu}(c)). The interface could thus be
periodically destabilized and reconstructed by the same process as in Ref.~\cite{Radulescu:2003}.
As a cause for the slip dynamics, one may invoke some
multi-valuedness of the shear rate inside the lubricating layers at a given shear stress,
which would lead to an instability near the walls similar to stick-slip or ``spurt'' effect
\cite{Dubbeldam:2003}. Further studies will focus on trying to isolate
the bulk behavior from that of sliding layers and determine how the two are coupled.

In summary, ultrasonic velocimetry allows the study of
sheared complex fluids on time scales of about 1~s. Such a
time resolution is crucial since a wrong picture of the flow may be deduced from
velocity profiles averaged on longer times with slower techniques.
In the case of a wormlike micelle solution,
various dynamical regimes were unveiled: the slow nucleation and
growth of a shear band as well as more complex faster behaviors.
These results, together with evidence for strong wall slip dynamics, 
demonstrate the importance of spatio-temporal fluctuations in 
shear-banded flows. More generally, they emphasize the need for time-resolved
measurements as well as dynamical theoretical approaches
in the study of complex fluid flows.

\begin{acknowledgments}
The authors wish to thank D. Anache and the ``Cellule Instrumentation'' of CRPP for technical help
with the experiment.
Fruitful discussions with A. Aradian, F. Molino, P. Olmsted, G. Porte, and J.-B. Salmon
are acknowledged.
\end{acknowledgments}


\begin{thebibliography}{30}
\bibitem{Remark:tempo}
A mechanical effect, such as a small amount of wobble, would lead to a single
time scale corresponding to the rotation period 0.75~s of the rotor. Moreover, in the system under study,
$\gammap=200$~s$^{-1}$ is far from any hydrodynamic or elastic instability.
\end{thebibliography}
\end{document}